\documentclass[11pt, oneside]{article}   	%
\usepackage{geometry}                		%
\usepackage{graphicx}				%
\usepackage{amssymb}
\usepackage{physics}
\usepackage[colorlinks=false, urlcolor=blue, linkcolor=red]{hyperref}
\usepackage{cite}
\usepackage[symbol]{footmisc}
\allowdisplaybreaks

\title{\bf Analytic solutions for the longitudinal and the transverse components of the 
vector potential in the Lorenz gauge}
\author
{Kuo-Ho Yang\\
Department of Engineering and Physics, St.~Ambrose University, Davenport, IA 52803 \\
E-mail: yangkuoho@sau.edu\vspace{2mm}\\
Robert D. Nevels\\
Department of Electrical and Computer Engineering, Texas A\ \&\ M University,\\
College Station, TX 77843\\
E-mail: r-nevels@tamu.edu
}
\date{}
\begin{document}
\maketitle
\begin{abstract}
We derive analytic solutions for the longitudinal and the transverse components of the vector potential in the Lorenz gauge for an arbitrary time-dependent charge-current distribution.
\end{abstract}\vspace{3mm}

In a recent note \cite{Hnizdo-2025}, Hnizdo pointed out a minor mis-statement in Sec. II of Jackson's paper \cite{Jackson-2002} where the longitudinal and the transverse components of the vector potential in the Lorenz gauge, ${\bf A}^{(L)}_\ell$ and ${\bf A}^{(L)}_{tr}$, were discussed.  This motivated us to look deeper into the problem to seek a better understanding of a mystery surrounding these vector components.  A previous analysis indicated that the longitudinal component ${\bf A}^{(L)}_\ell$ contained a term propagating with a speed greater than $c$ from physical charge and current densities (Sec.~II.C of Ref.~\cite{Yang-2005}).   In this note, we show that the mathematics developed to solve potentials from Maxwell's equations for potentials \cite{Yang-Nevels-2025-1, Yang-Nevels-2025-2} can be applied to derive analytic solutions for ${\bf A}^{(L)}_\ell$ and ${\bf A}^{(L)}_{tr}$.  Our solutions are expressed in familiar quantities in electrodynamics and are valid for an arbitrary time-dependent charge-current distribution.

We consider localized charge and current densities, $\rho ({\bf r}, t)$ and ${\bf J} ({\bf r}, t)$, which are turned on at $t_0$.  The electric field ${\bf E}$, the magnetic field ${\bf B}$, and Maxwell's equations for potentials ${\bf A}$ and $\Phi$ are (in Gaussian units):
\begin{equation}
{\bf E} = - \grad \Phi - {1 \over c} {\partial {\bf A} \over \partial t},
\qquad \qquad
{\bf B} = \grad \times {\bf A},
\label{eq-A1}
\end{equation}
\begin{equation}
\grad^2 \Phi + {1 \over c} {\partial \over \partial t} ( \grad \cdot {\bf A} )
= - 4 \pi \rho,
\label{eq-A2}
\end{equation}
\begin{equation}
\left( \grad^2 - {1 \over c^2} {\partial^2 \over \partial t^2}  \right) {\bf A}  
= - { 4 \pi \over c } {\bf J} 
  + \grad \left( \grad \cdot {\bf A} + {1 \over c} {\partial \Phi \over \partial t} \right).
\label{eq-A3}
\end{equation}

To discuss the longitudinal and the transverse components of the Lorenz-gauge vector potential, we start with the Coulomb gauge:
\begin{equation}
 \grad \cdot {\bf A}^{(C)} = 0.
 \label{eq-A4}
\end{equation}
Thus, eqs.~(\ref{eq-A2})-(\ref{eq-A3}) become:
\begin{equation}
\grad^2 \Phi^{(C)} = - 4\pi \rho,
\label{eq-A5}
\end{equation}
\begin{equation}
\left( \grad^2 - {1 \over c^2} {\partial^2 \over \partial t^2} \right) {\bf A}^{(C)}
= - {4 \pi \over c} ({\bf J} - {\bf J}_\ell)
= - {4 \pi \over c} {\bf J}_{tr}.
\label{eq-A6}
\end{equation}
Here the transverse current density ${\bf J}_{tr}$ is related to the current density ${\bf J}$ and the longitudinal current density ${\bf J}_\ell$ by (see, {\it e.g.}, Ref.~\cite{Jackson-2002}):
\begin{equation}
{\bf J}_\ell 
= - {1 \over 4\pi} \grad \int { \grad' \cdot {\bf J} ({\bf r}', t) \over |{\bf r} - {\bf r}'| } d^3r'
= {1 \over 4\pi} \grad {{\partial \over \partial t}} \int { \rho({\bf r}', t) \over |{\bf r} - {\bf r}'| } d^3r'
= {1 \over 4\pi} \grad {{\partial \over \partial t} \Phi^{(C)}},
\label{eq-A7}
\end{equation}
\begin{equation}
{\bf  J}_{tr} = {\bf  J} - {\bf  J}_\ell 
= {1 \over 4 \pi} \grad \times \left( \grad \times \int {{\bf J}({\bf r}',t) \over |{\bf r} - {\bf r}'|}d^3r'  \right) 
= {c \over 4 \pi} \grad \times \left( \grad \times {\bf A}_\infty  \right),
\label{eq-A8}
\end{equation}
\begin{equation}
{\bf A}_\infty = {1 \over c} \int {{\bf J}({\bf r}', t) \over |{\bf r} - {\bf r}'|} d^3r',
\qquad \qquad \qquad
\grad^2 {\bf A}_\infty = - {4 \pi \over c} {\bf J},
\label{eq-A9}
\end{equation}
where the subscript $\infty$ indicates that ${\bf A}_\infty$ propagates instantaneously from the current density ${\bf J}$. 

We now consider the Lorenz gauge, 
\begin{equation}
\grad \cdot {\bf A}^{(L)} + {1 \over c} {\partial \over \partial t} \Phi^{(L)} = 0.
\label{eq-A10}
\end{equation}
The equations for the scalar and the vector potentials are:
\begin{equation}
\left( \grad^2 - {1 \over c^2} {\partial^2 \over \partial t^2} \right) \left( \Phi^{(L)}, {\bf A}^{(L)} \right)
= \left( - 4 \pi \rho, -{4\pi \over c}{\bf J} \right).
\label{eq-A11}
\end{equation}
If we use ${\bf J} = {\bf J}_\ell + {\bf J}_{tr}$ in eq.~(\ref{eq-A11}), the longitudinal and the transverse components of the Lorenz-gauge vector potential satisfy the equations:
\begin{equation}
\left( \grad^2 - {1 \over c^2} {\partial^2 \over \partial t^2} \right) {\bf A}^{(L)}_\ell
= - {4\pi \over c}{\bf J}_\ell 
= - {1 \over c} \grad {{\partial \over \partial t}} \Phi^{(C)},
\label{eq-A12}
\end{equation}
\begin{equation}
\left( \grad^2 - {1 \over c^2} {\partial^2 \over \partial t^2} \right)  {\bf A}^{(L)}_{tr}
= - {4\pi \over c}{\bf J}_{tr}
= - \grad \times ( \grad \times {\bf A}_\infty).
\label{eq-A13}
\end{equation}

We note that there is no need to solve for {\it both} ${\bf A}^{(L)}_\ell$ and ${\bf A}^{(L)}_{tr}$ from eq.~(\ref{eq-A12})-(\ref{eq-A13}).  It is because ${\bf A}^{(L)}_\ell + {\bf A}^{(L)}_{tr} = {\bf A}^{(L)}$.  If we solve for ${\bf A}^{(L)}$ and one of the components, we can use ${\bf A}^{(L)}_{tr} = {\bf A}^{(L)} - {\bf A}^{(L)}_\ell$ or ${\bf A}^{(L)}_\ell = {\bf A}^{(L)} - {\bf A}^{(L)}_{tr}$ to derive the other component.   In this note, we do solve both components to show that our mathematics can handle the challenges.

\medskip
\begin{flushleft}
{\bf Method 1: Vector potential in the Coulomb gauge}
\end{flushleft}

In Ref.~\cite{Yang-Nevels-2025-1}, it was shown that the vector potential ${\bf A}$ in an arbitrary gauge has the solution:
\begin{equation}
{\bf A} ({\bf r}, t) = {\bf A}^{(L)} ({\bf r}, t) + c \grad \int \left[ \Phi^{(L)} ({\bf r}, t) - \Phi({\bf r}, t) \right] dt,
\label{eq-A14}
\end{equation}
where ${\Phi}$ is the scalar potential in the same gauge.

From eqs.~(\ref{eq-A6}), (\ref{eq-A13}) and (\ref{eq-A14}), We see that the transverse component of the vector potential in the Lorenz gauge has the expression:
\begin{equation}
{\bf A}^{(L)}_{tr} = {\bf A}^{(C)} = {\bf A}^{(L)} + c \grad \int \left( \Phi^{(L)} - \Phi^{(C)} \right) dt.
\label{eq-A15}
\end{equation}
Therefore, the longitudinal component of the Lorenz-gauge vector potential is:
\begin{equation}
{\bf A}^{(L)}_\ell  = {\bf A}^{(L)} - {\bf A}^{(L)}_{tr} = c \grad \int \left( \Phi^{(C)} - \Phi^{(L)} \right) dt.
\label{eq-A16}
\end{equation}
Below is a proof that ${\bf A}^{(L)}_\ell$ in eq.~(\ref{eq-A16}) is indeed a valid solution of eq.~(\ref{eq-A12}).  We apply $(\grad^2 - c^{-2} \partial^2/\partial t^2)$ to eq.~(\ref{eq-A16}) and use (\ref{eq-A5})-(\ref{eq-A11}) to get (see Ref.~\cite{Yang-Nevels-2025-2}):
\begin{eqnarray}
\left(\grad^2 - {1 \over c^2}{\partial^2 \over \partial t^2} \right) {\bf A}^{(L)}_\ell
= c \grad \int  \left[ \left(\grad^2 \Phi^{(C)}  - {1 \over c^2}{\partial^2 \over \partial t^2} \Phi^{(C)} \right)
        -  \left(\grad^2 - {1 \over c^2}{\partial^2 \over \partial t^2} \right) \Phi^{(L)} \right] dt
 \nonumber
 \\
 = c \grad \int \left[ \left( - 4\pi \rho -   {1 \over c^2}{\partial^2 \over \partial t^2}\Phi^{(C)}  \right)
  -  \left( - 4\pi {\rho} \right) \right] dt
= - {1 \over c} \grad {{\partial \over \partial t}} \Phi^{(C)}
= - {4\pi \over c} {\bf J}_\ell.
\qquad \quad \quad
\label{eq-A17}
\end{eqnarray}

\medskip
\begin{flushleft}
{\bf Method 2: Jackson's equation for the longitudinal component}
\end{flushleft}

To solve ${\bf A}^{(L)}_\ell$ directly from eq.~(\ref{eq-A12}), we write ${\bf A}^{(L)}_\ell = \grad \Psi$ and use it in eq.~(\ref{eq-A12}) to derive an equation for $\Psi$ (in Jackson's eq.~(2.10) in Ref.~\cite{Jackson-2002}):
\begin{equation}
\left(\grad^2 - {1 \over c^2}{\partial^2 \over \partial t^2} \right) \Psi
= - {1 \over c} \left( {\partial \over \partial t} \Phi^{(C)} \right).
\label{eq-A18}
\end{equation}
To solve $\Psi$, we differentiate both sides by $t$ (see Ref.~\cite{Yang-Nevels-2025-2}):
\begin{equation}
\left( \grad^2 - {1 \over c^2} {\partial^2 \over \partial t^2} \right) {\partial \Psi \over \partial t}
= - {1 \over c^2}{\partial^2 \over \partial t^2} \left(c \Phi^{(C)} \right)
= - \grad^2 \left(c \Phi^{(C)} \right) 
+ \left( \grad^2 -  {1 \over c^2}{\partial^2 \over \partial t^2} \right)\left(c \Phi^{(C)} \right).
\label{eq-A19}
\end{equation}
We move the last term to the LHS and use eq.~(\ref{eq-A5}) to get
\begin{equation}
\left( \grad^2 - {1 \over c^2} {\partial^2 \over \partial t^2} \right)
   \left( {\partial \Psi \over \partial t} - c\Phi^{(C)} \right)
=  - \grad^2 \left(c \Phi^{(C)} \right) 
= 4\pi c \rho.
\label{eq-A20}
\end{equation}
This wave equation can be solved by the $c$-retarded propagation method:
\begin{equation}
 {\partial \Psi  \over \partial t} - c\Phi^{(C)} 
 = - c \int {\rho({\bf r}', t' =  t-|{\bf r} - {\bf r}'|/c) \over |{\bf r} - {\bf r}'|} d^3r'
 = - c \Phi^{(L)},
\label{eq-A21}
\end{equation}
\begin{equation}
\Psi = c \int \left( \Phi^{(C)} - \Phi^{(L)} \right) dt,
\qquad \qquad 
{\bf A}^{(L)}_\ell = \grad \Psi = c \grad \int \left( \Phi^{(C)}  - \Phi^{(L)} \right) dt.
\label{eq-A22}
\end{equation}
It is obvious that the longitudinal component ${\bf A}^{(L)}_\ell$ in eq.~(\ref{eq-A22}) and the associated transverse component ${\bf A}^{(L)}_{tr} = {\bf A}^{(L)} - {\bf A}^{(L)}_\ell$ are in total agreement with the results in eqs.~(\ref{eq-A15})-(\ref{eq-A16}).

\medskip
\begin{flushleft}
{\bf Method 3: Jackson's equation for the transverse component}
\end{flushleft}

To solve for the transverse component we write ${\bf A}^{(L)}_{tr} = \grad \times {\bf V}$ and use it in eq.~(\ref{eq-A13}) to derive an equation for ${\bf V}$ (in Jackson's equation (2.10) in Ref.~\cite{Jackson-2002}):
\begin{equation}
\left( \grad^2 - {1 \over c^2} {\partial^2 \over \partial t^2} \right) {\bf V} = - \grad \times {\bf A}_{\infty},
\label{eq-A23}
\end{equation}
where ${\bf A}_\infty$ is defined in eq.~(\ref{eq-A9}).  To solve ${\bf V}$, we write ${\bf V} = \grad \times {\bf W}$ and solve ${\bf W}$ from the equation:
\begin{equation}
\left( \grad^2 - {1 \over c^2} {\partial^2 \over \partial t^2} \right) {\bf W} = - {\bf A}_{\infty}.
\label{eq-A24}
\end{equation}
We do $\partial^2/\partial t^2$ to both sides of the equation to get
\begin{eqnarray}
\left( \grad^2 - {1 \over c^2} {\partial^2 \over \partial t^2} \right) {\partial^2 {\bf W} \over \partial t^2} 
= - {1 \over c^2} {\partial^2 \over \partial t^2} \left(c^2{\bf A}_\infty \right)
= - \grad^2  \left(c^2{\bf A}_\infty \right)
   + \left( \grad^2 - {1 \over c^2} {\partial^2 \over \partial t^2} \right) \left(c^2{\bf A}_\infty \right).
\label{eq-A25}
\end{eqnarray}
We move the last term to the LHS and use eq.~(\ref{eq-A9}) to get
\begin{equation}
\left( \grad^2 - {1 \over c^2} {\partial^2 \over \partial t^2} \right)
\left( {\partial^2 {\bf W} \over \partial t^2} - c^2 {\bf A}_\infty \right)
= - c^2 \grad^2 {\bf A}_\infty
= {4\pi c} {\bf J}.
\label{eq-A26}
\end{equation}
Hence, 
\begin{equation}
{\partial^2 {\bf W} \over \partial t^2} - c^2 {\bf A}_\infty 
= - c \int {{\bf J}({\bf r}', t' = t - |{\bf r} - {\bf r}'| / c) \over |{\bf r} - {\bf r}'|} d^3r'
= - c^2 {\bf A}^{(L)},
\qquad \qquad \qquad  
\label{eq-A27}
\end{equation}
\begin{equation}
{\partial^2 {\bf W} \over \partial t^2} = c^2 {\bf A}_\infty - c^2 {\bf A}^{(L)},
\qquad \qquad \qquad \qquad \qquad \qquad \qquad \qquad \qquad \qquad \qquad \ \,
\label{eq-A28}
\end{equation}
\begin{eqnarray}
{\partial^2 {\bf A}^{(L)}_{tr}  \over \partial t^2} = {\partial^2 ( \grad \times {\bf V} ) \over \partial t^2}
= {\partial^2 [ \grad \times ( \grad \times {\bf W} ) ] \over \partial t^2} 
= \grad \times \left( \grad \times {\partial^2 {\bf W} \over \partial t^2} \right)
\qquad \qquad \qquad 
\nonumber
\\
= c^2 \left[ \grad \times \left( \grad \times {\bf A}_\infty \right) \right]  
- c^2 \left[  \grad \times ( \grad \times {\bf A}^{(L)} ) \right]
\qquad \qquad \qquad \qquad \qquad \quad \ \,
\nonumber
\\
= c^2 \left[\grad \left(\grad \cdot {\bf A}_\infty \right)  - \grad^2 {\bf A}_\infty \right] 
 - c^2 \left [\grad (\grad \cdot {\bf A}^{(L)})  - \grad^2 {\bf A}^{(L)} \right] 
\qquad \qquad \ \ \ \,
\nonumber
\\
= c^2 \left( -{1 \over c} \grad {{\partial \over \partial t}} \Phi^{(C)} + {4\pi \over c}{\bf J} \right)
- c^2 \left( -{1 \over c} \grad {{\partial \over \partial t}} \Phi^{(L)} 
    + {4\pi \over c}{\bf J} - {1 \over c^2} {\partial^2 \over \partial t^2} {\bf A}^{(L)} \right)
\, \ \,
\nonumber
\\
=  {\partial^2 \over \partial t^2} {\bf A}^{(L)} + c \grad {{\partial \over \partial t}} \left( \Phi^{(L)} - \Phi^{(C)} \right).
\qquad \qquad \qquad \qquad \qquad \qquad \qquad \quad \ \ \ \,
\label{eq-A29}
\end{eqnarray}
It is obvious that eq.~(\ref{eq-A29}) totally agrees with the result for ${\bf A}^{(L)}_{tr}$ in eq.~(\ref{eq-A15}).

\medskip
\begin{flushleft}
{\bf Method 4: Green's function method}
\end{flushleft}

A different way to solve eqs.~(\ref{eq-A18}) and (\ref{eq-A24}) is to do straightforward integrations using the $c$-retarded and the instantaneous Green's functions, $G({\bf r}, t | c |{\bf r}', t')$ and $G({\bf r}, t | \infty |{\bf r}', t')$.  Refer to the Appendix for the definitions of symbols and the important result in eq.~(\ref{eq-ApA6}) for $G({\bf r}, t |c|\infty|{\bf r}', t')$.  From eq.~(\ref{eq-A18}), we obtain $\Psi$ as follows:
\begin{eqnarray}
\Psi({\bf r}, t) = {1 \over 4\pi c} {\partial \over \partial t} \int d^3r''dt'' G({\bf r}, t | c |{\bf r}'', t'') 
     \Phi^{(C)}({\bf r}'', t'')
\qquad \qquad \qquad \qquad \quad \ 
\nonumber
\\
= {1 \over 4\pi c} {\partial \over \partial t}  \int d^3r''dt'' G({\bf r}, t | c |{\bf r}'', t'') 
                   \int G({\bf r}'', t'' | \infty |{\bf r}', t') \rho({\bf r}',t ') d^3r'dt'
 \nonumber
 \\
 = {1 \over 4\pi c} {\partial \over \partial t} \int G({\bf r}, t |c|\infty|{\bf r}', t') \rho({\bf r}', t') d^3r'dt'
\qquad \qquad \qquad \qquad \qquad \ \ \,
 \nonumber
 \\
 = c \int dt \int \left[ G({\bf r}, t |\infty| {\bf r}', t') - G({\bf r}, t |c| {\bf r}', t') \right] \rho({\bf r}', t') d^3r'dt'
 \qquad  \qquad 
  \nonumber
 \\
 = c \int \left[ \Phi^{(C)}({\bf r}, t) - \Phi^{(L)}({\bf r}, t) \right]dt,
  \qquad  \qquad  \qquad  \qquad  \qquad  \qquad  \qquad \ \ 
\label{eq-A30}
\end{eqnarray}
which agrees eq.~(\ref{eq-A22}).
The solution for ${\bf W}$ in eq.~(\ref{eq-A24}) is derived similarly:
\begin{eqnarray}
{\bf W}({\bf r}, t) = {1 \over 4\pi} \int d^3r''dt'' G({\bf r}, t | c |{\bf r}'', t'') {\bf A}_\infty ({\bf r}'', t'')
\qquad \qquad \qquad \qquad \qquad \quad \ \, 
\nonumber
\\
= {1 \over 4\pi c} \int d^3r''dt'' G({\bf r}, t | c |{\bf r}'', t'') 
                   \int G({\bf r}'', t'' | \infty |{\bf r}', t') {\bf J}({\bf r}',t ') d^3r'dt'
 \quad \ 
 \nonumber
 \\
 = {1 \over 4\pi c} \int G({\bf r}, t | c | \infty |{\bf r}', t') {\bf J}({\bf r}', t') d^3r'dt'
\qquad \qquad \qquad \qquad \qquad \qquad \, 
 \nonumber
 \\
 = c \int dt \int dt \int \left[ G({\bf r}, t |\infty| {\bf r}', t') - G({\bf r}, t |c| {\bf r}', t') \right] {\bf J}({\bf r}', t') d^3r'dt'
 \qquad 
  \nonumber
 \\
 = c^2  \int dt \int dt \left[ {\bf A}_\infty({\bf r}, t) - {\bf A}^{(L)}({\bf r}, t) \right],
  \qquad  \qquad  \qquad  \qquad  \qquad  \qquad  \ 
\label{eq-A31}
\end{eqnarray}
which agrees with eq.~{\ref{eq-A28}).

In conclusion, we have derived analytic solutions of the longitudinal and transverse components, ${\bf A}^{(L)}_\ell$ and ${\bf A}^{(L)}_{tr}$, of the Lorenz-gauge vector potentials using different methods and arriving at the same results.  In particular, we have derived exact solutions for Jackson's original equations for the ${\bf A}^{(L)}_\ell$ and ${\bf A}^{(L)}_{tr}$ in eqs.~(2.9)-(2.10) of Ref.~\cite{Jackson-2002}.

\appendix 
\section{Appendix: Mathematical properties of propagating Green's functions}
\setcounter{equation}{0}
\renewcommand{\theequation}{A.\arabic{equation}}

For a more detailed discussion of the mathematics in this Appendix and its applications, refer to Ref.~\cite{Yang-Nevels-2025-3}.  We first define the $v$-propagating Green's function by
\begin{equation}
G({\bf r}, t |v| {\bf r}', t') = { \delta(t - {|{\bf r} - {\bf r}'| \over v} - t') \over  |{\bf r} - {\bf r}'|},
\label{eq-ApA1}
\end{equation}
\begin{equation}
\left( \grad^2 - {1 \over v^2} {\partial^2 \over \partial t^2} \right) G({\bf r}, t |v| {\bf r}', t') 
= - 4 \pi \delta({\bf r} - {\bf r}') \delta(t - t').
\label{eq-ApA2}
\end{equation}
Unless explicitly stated, we normally only consider the retarded solutions with $v > 0$.

We now define a two-speed propagating Green's function with speeds $c$ and $v$ by\cite{Yang-2005}:
\begin{equation}
G({\bf r}, t |c |v| {\bf r}', t') = \int G({\bf r}, t |c| {\bf r}'', t'')  G({\bf r}'', t'' |v| {\bf r}', t') d^3r''dt'',
\label{eq-ApA3}
\end{equation}
\begin{equation}
\left( \grad^2 - {1 \over v^2} {\partial^2 \over \partial t^2} \right) 
\left( \grad^2 - {1 \over c^2} {\partial^2 \over \partial t^2} \right)
G({\bf r}, t |c |v| {\bf r}, t') 
= (-4\pi)^2 \delta({\bf r} - {\bf r}') \delta(t - t').
\label{eq-ApA4}
\end{equation}
We can derive the following result (see eq.~(14) of Ref.~\cite{Yang-Nevels-2025-3}):
\begin{equation}
{\partial^2 \over \partial t^2} \left[ {v^2 - c^2 \over 4\pi v^2} G({\bf r}, t |c |v| {\bf r}', t') \right]
= c^2 \left[ G({\bf r}, t |v| {\bf r}', t') - G({\bf r}, t |c | {\bf r}', t') \right].
\label{eq-ApA5}
\end{equation}
In the limit $v \to \infty$, the above result becomes
\begin{equation}
 G({\bf r}, t |c |\infty| {\bf r}', t') 
 = 4 \pi c^2 \int dt \int dt \left[ G({\bf r}, t |\infty| {\bf r}', t') - G({\bf r}, t |c| {\bf r}', t') \right].
 \label{eq-ApA6}
\end{equation}

\end{document}